\documentclass[aps,prd,superscriptaddress,showpacs,amsmath,amssymb]{revtex4}
\usepackage{graphicx}
\usepackage{epsf}

\usepackage[]{latexsym}
\usepackage{bm}

\newcommand{\be}{\begin{equation}}\newcommand{\ee}{\end{equation}}
\newcommand{\bea}{\begin{eqnarray}}\newcommand{\eea}{\end{eqnarray}}
\newcommand{\brr}{\begin{array}}\newcommand{\err}{\end{array}}
\newcommand{\bit}{\begin{itemize}}\newcommand{\eit}{\end{itemize}}
\newcommand{\ben}{\begin{enumerate}}\newcommand{\een}{\end{enumerate}}

\newcommand{\ba}{\begin{array}}
\newcommand{\ea}{\end{array}}

\def\lan{\langle}
\def\lf{\left}

\def\non{\nonumber}\def\ran{\rangle}

\def\ri{\right}
\def\al{\alpha}
\def\de{\delta}
\def\te{\theta}
\def\la{\lambda}\def\si{\sigma}
\def\om{\omega}

\def\1{{_{1}}}\def\2{{_{2}}}

\newcommand{\ide}{1\hspace{-1mm}{\rm I}}

\def\noHe0{:\;\!\!\;\!\!:H_e(0):\;\!\!\;\!\!:}
\def\noHm0{:\;\!\!\;\!\!:H_\mu(0):\;\!\!\;\!\!:}
\def\nof{:\;\!\!\;\!\!:}
\def\vect#1{{\bm #1}}

\def\lan{\langle}
\def\lf{\left}

\def\non{\nonumber}
\def\ran{\rangle}

\def\ri{\right}

\def\al{\alpha}
\def\de{\delta}
\def\te{\theta}
\def\la{\lambda}
\def\si{\sigma}
\def\om{\omega}

\def\1{{_{1}}}\def\2{{_{2}}}
\def\nof{:\;\!\!\;\!\!:}

\def\wwQ{Q}
\begin{document}

\title{Non--abelian gauge structure in neutrino mixing}

\author{Massimo Blasone}
\affiliation{DMI,
Universit\`a degli Studi di Salerno, Via Ponte don Melillo,
I-84084 Fisciano (SA), Italy} \affiliation{INFN Sezione di Napoli,
Gruppo collegato di Salerno, Italy}

\author{Marco Di Mauro}
\affiliation{DMI,
Universit\`a degli Studi di Salerno, Via Ponte don Melillo,
I-84084 Fisciano (SA), Italy} \affiliation{INFN Sezione di Napoli,
Gruppo collegato di Salerno, Italy}

\author{Giuseppe Vitiello}
\affiliation{DMI,
Universit\`a degli Studi di Salerno, Via Ponte don Melillo,
I-84084 Fisciano (SA), Italy} \affiliation{INFN Sezione di Napoli,
Gruppo collegato di Salerno, Italy}


\begin{abstract}
We discuss the existence of  a non--abelian gauge
structure associated with 
flavor mixing. In the specific case of two flavor
mixing of Dirac neutrino fields, we  show that this reformulation 
allows to define flavor neutrino states
which preserve the Poincar\'e structure. 
Phenomenological consequences of
our analysis are explored.
\end{abstract}

\maketitle

\section{Introduction}

The study of neutrino mixing and oscillations is of
utmost importance in contemporary theoretical and
experimental physics, due to the experimental discovery of
neutrino oscillations.
 At a theoretical level, one important issue is the one of a correct
 definition of the  flavor states i.e. the ones describing
oscillating neutrinos.
In the standard quantum mechanical treatment, the well known Pontecorvo
states \cite{Bilenky:1978nj} are used and oscillation formulas are derived,
which can describe efficiently the main aspects of such a phenomenon.
However, it is clear that Pontecorvo states are only approximate since
they are not eigenstates of the
flavor neutrino charges.
Thus Pontecorvo states lead to violation of the conservation of leptonic charge
in the neutrino production vertices \cite{BCTV2005,Nishi:2008sc}.

The solution to the above problem has been found
in the context of Quantum
Field Theory (QFT). Indeed, by considering mixing at level of fields, rather
than postulating it as a property of states, unexpected features
emerged \cite{BV95}.
It has been found that field mixing is associated with
inequivalent representation of the canonical anticommutation relations,
i.e. the vacuum for the mass
eigenstates of neutrinos has been found to  be unitarily inequivalent
to the vacuum for the flavor
eigenstates of neutrinos -- the flavor vacuum.
The nonperturbative vacuum structure associated with field
mixing has been found to be a very general feature, independently
of the nature of the fields \cite{Ji,Fujii:1999xa,bosonmix,3flav,Blasone:2003hh}.
It has also  been shown that
it leads to modifications of the flavor oscillation formulae \cite{BHV98,Blasone:2002wp,Ji,Fujii:1999xa}.

In QFT flavor states can be straightforwardly defined
as eigenstates of the
flavor charges which are derived in a canonical way
from the symmetry properties of the neutrino Lagrangian \cite{BJV01}.
It has been shown that
states defined in this way restore flavor charge conservation
in weak interaction vertices, at tree level \cite{Blasone:2006jx}. Moreover,
such states turn out to be eigenstates of the momentum operator.

Despite the above mentioned results, the QFT treatment of flavor states
still presents some open problems. One such issue is Lorentz invariance. Indeed,
the flavor vacuum is not Lorentz invariant being explicitly time-dependent.
As a consequence, flavor states cannot be interpreted in terms of irreducible representations of the Poincar\'{e} group. A possible way to recover
Poincar\'{e} invariance for mixed fields  has been explored in Refs.\cite{BlaMagueijo} where  non--standard dispersion relations
for the mixed particles have been related to  nonlinear realizations of the Poincar\'{e} group \cite{MagSmol}. Another interesting issue concerns the invariance of the flavor oscillation formulas under Lorentz boosts \cite{Blasone:2005tm}.

The relation of neutrino masses and mixing with a possible violation of the Poincar\'{e} and $CPT$
symmetries has been the subject of many efforts in the last decade \cite{Pakvasa:2001yw}. A related line
of research concerns the use of neutrino mixing and oscillations as a sensitive probe
for quantum gravity effects, as quantum gravity induced decoherence is expected to affect neutrino oscillations \cite{AmelinoCamelia:2007kx}. Such effects have also been connected \cite{Mavromatos:2007ak} to the non trivial structure of
the flavor vacuum introduced in  \cite{BV95}.

In this paper we propose an approach to the mixing of particles which
overcomes the problems mentioned above. The basic idea is to
view the mixing phenomenon as the result of the interaction of the neutrino
fields with an external field, which as
we shall see appears to be a non-abelian gauge field.
This point of view allows to treat formally the mixed fields as free fields,
avoiding in this way the problems with
their interpretation in terms of the Poincar\'{e} group.
The  violation of relativistic invariance
is now seen as a consequence of the presence of a fixed external field,
which defines a preferred direction in spacetime.

Our approach  enables us to define flavor neutrino states
which are simultaneous eigenstates of the flavor charges, of
the momentum operators and of a new Hamiltonian operator for
the mixed fields whose definition  naturally emerges
from our approach. This operator can be interpreted as
the energy which can be extracted from flavor neutrinos through
scattering. We discuss a possible test for our theoretical scheme,
by looking at mixed neutrinos in the $\beta$ decay, where the
endpoint of the electron energy spectrum turns out to be different in our approach
with respect to the standard prediction.

In the present paper, we  consider only the mixing of two Dirac fermion fields.
Similar results  hold also for the case of mixing of boson fields, and for  the case of three flavors. An analysis of these
instances will be presented elsewhere.

\section{Two-flavor neutrino mixing}

We begin with the Lagrangian density describing two mixed neutrino
fields:
\bea\label{Lagrflav} {\cal L} &=& {\bar \nu}_e \lf( i
\not\!\partial -
  m_{e}\ri)\nu_e  +  {\bar \nu}_\mu \lf( i \not\!\partial -
  m_{\mu}\ri)\nu_\mu \,- \, m_{e \mu} \,\lf({\bar \nu}_{e}\,
\nu_{\mu} \,+\, {\bar \nu}_{\mu} \,\nu_{e}\ri). \eea
The standard treatment of the problem is based on the observation
that this Lagrangian, being quadratic, can be diagonalized by a
canonical transformation of the field operators (called the \emph{mixing transformation}):
\bea \label{PontecorvoMix}
\nu_e &=& \nu_1\cos\theta + \nu_2\sin\theta\\[1mm]
\nu_{\mu} &=& -\nu_1\sin\theta + \nu_2\cos\theta, \eea
 so that one simply gets
the sum of two free Dirac Lagrangians:
\bea\label{Lagrmass} {\cal L} &=& {\bar \nu}_1 \lf( i
\not\!\partial -
  m_{1}\ri)\nu_1  +  {\bar \nu}_2 \lf( i \not\!\partial -
  m_{2}\ri)\nu_2. \; \eea
In the above equations,  $\theta$ is the mixing angle and
$ m_{e} = m_{1}\cos^{2}\theta + m_{2} \sin^{2}\theta$,
$m_{\mu} = m_{1}\sin^{2}\theta + m_{2} \cos^{2}\theta$,
$m_{e\mu} =(m_{2}-m_{1})\sin\theta \cos\theta\,$.

From the above lagrangian,  one can derive the
canonical  energy-momentum tensor:
\bea \non
 T_{\rho\sigma}&=& {\bar \nu}_e i
\gamma_{\rho}\partial_{\sigma}\nu_e - \eta _{\rho\sigma}{\bar \nu}_e
(i\gamma^{\lambda}\partial_{\lambda} - m_e)\nu_e + {\bar \nu}_{\mu} i
\gamma_{\rho}\partial_{\sigma}\nu_{\mu} - \eta _{\rho\sigma}{\bar
\nu}_{\mu} (i\gamma^{\lambda}\partial_{\lambda} - m_{\mu})\nu_{\mu} + \eta_{\rho\sigma}m_{e\mu}({\bar \nu}_e \nu_{\mu} +  {\bar
\nu}_{\mu} \nu_e)
\\[1mm] \label{Temu}
&=& {\bar \nu}_1 i
\gamma_{\rho}\partial_{\sigma}\nu_1 - \eta _{\rho\sigma}{\bar \nu}_1
(i\gamma^{\lambda}\partial_{\lambda} - m_1)\nu_1 + {\bar \nu}_2 i
\gamma_{\rho}\partial_{\sigma}\nu_2- \eta _{\rho\sigma}{\bar
\nu}_2 (i\gamma^{\lambda}\partial_{\lambda} - m_2)\nu_2,\eea
where $\eta_{\rho\sigma}=\textrm{diag}(+1,-1,-1,-1)$ is the Minkowskian metric
tensor. From this tensor it follows the total Hamiltonian:
\bea \label{H12}
{H} &=& \int d^3 \mathbf{x}\,
{T}^{00}=  \int d^3 \mathbf{x}\, \nu_1^{\dagger}
\lf(-i\vect{\al}\cdot\vect{\nabla}+\beta
m_1\ri) \nu_1 + \int d^3 \mathbf{x}
\,\nu_2^{\dagger}\lf(-i\vect{\al}\cdot\vect{\nabla}+\beta m_2\ri)
\nu_2
.\eea
which is just the sum of the two free field Hamiltonians: $H=H_1+H_2$.

Analogously, a momentum operator is defined as:
\bea \label{Pfree}
{P}^i &=& \int d^3\mathbf{x} \,{T}^{0i}\,=\,
i \int d^3\mathbf{x}\,
\nu_1^{\dagger}\partial^i \nu_1 + i \int d^3\mathbf{x}\,
\nu_2^{\dagger}\partial^i \nu_2
,\qquad i=1,2,3 \eea
which again is the sum of two free field contributions.

To the  free fields $\nu_i$ there are associated two conserved (Noether)
charges:
\bea\label{su2noether}
&&Q_j\, = \,\int d^{3}{\bf x} \,  \nu_{j}^{\dag}(x)\;\nu_{j}(x)\,,
\qquad j=1,2,
\eea
with the total charge $Q=Q_1+Q_2$.
The analysis of symmetries of the Lagrangian in the flavor basis
Eq.(\ref{Lagrflav}), leads to the identification of the (non conserved) flavor
charges \cite{BJV01}:
\bea \label{flavcharges} && Q_{\sigma}(x_0) \,= \,\int d^{3}{\bf
x}\,
 \nu_{\sigma}^{\dag}(x)\;\nu_{\sigma}(x)~,
\quad \sigma=e,\mu,
\eea
with $Q_{e}(x_0) + Q_{\mu}(x_0) = Q$.  Flavor charges describe
the phenomenon of neutrino oscillations, see Appendix A.

It is interesting to consider the relation among the two sets of charges:
\bea\label{carichemix1} Q_{e}(x_0) &=& \cos^2\te\;  Q_{1} +
\sin^2\te \; Q_{2} + \sin\te\cos\te \int d^3{\bf x} \lf[\nu_1^\dag
(x) \nu_2(x) + \nu_2^\dag(x) \nu_1(x)\ri]\,,
\\
\label{carichemix2} Q_{\mu}(x_0)&=& \sin^{2}\te \; Q_{1}
+\cos^{2}\te \; Q_{2} - \sin\te \cos\te \int d^3{\bf x}
\lf[\nu_1^\dag(x) \nu_2(x) + \nu_2^\dag(x) \nu_1(x)\ri]\,. \eea
The appearance of  terms that cannot be written
in terms of $Q_j$ is related to a non-trivial structure of the
flavor Hilbert space \cite{BV95}, see Appendix A.

\section{Flavor mixing as a non-abelian gauge theory}

We now show that the
Lagrangian Eq.(\ref{Lagrflav}) can be formally
written  as a non-abelian gauge theory. In the following we shall use the conventions of Ref. \cite{LeiteLopes:1981fr}.

The starting point is the observation that the mixing interaction
can be consistently viewed as the interaction of the flavor
neutrino fields with a constant external gauge field. The most direct way of seeing
this goes through the Euler--Lagrange equations corresponding to
the Lagrangian (\ref{Lagrflav}), namely:
\bea i \partial_0 \nu_e &=& (-i \vect{\al}\cdot\vect{\nabla} + \beta m_e)\nu_e
+ \beta m_{e\mu} \nu_{\mu} \\[2mm] i \partial_0 \nu_{\mu} &=& (-i
\vect{\al}\cdot\vect{\nabla} + \beta m_{\mu})\nu_{\mu} + \beta m_{e\mu}
\nu_e,\eea
where $\alpha_i$, $i=1,2,3$ and $\beta$ are the usual Dirac
matrices in a given representation. Here we choose the following representation:
\bea \alpha_i=\lf(\ba{cc}0&\sigma_i\\\sigma_i&0\ea\ri),\qquad
\beta= \lf(\ba{cc}\ide&0\\0&-\ide\ea\ri),\eea
where $\sigma_i$ are the Pauli
matrices and $\ide$ is the $2\times 2$ identity matrix.
The Euler--Lagrange  equations  can be compactly written as
follows:
\bea iD_0 \nu_f= (-i\vect{\al}\cdot\vect{\nabla} + \beta M_d)\nu_f,\eea
where $\nu_f= (\nu_e, \nu_\mu )^T$ is the flavor doublet and
$M_d={\rm diag}(m_e,m_{\mu})$ is a diagonal mass matrix. We have
defined the (non-abelian) covariant derivative:
\bea \label{covdevmix}
D_0 :=\partial _0 + i\, m_{e\mu}\,\beta\,\sigma_1,  \eea
%
where  $m_{e\mu}=\frac{1}{2}\,\tan 2\theta \,\delta m$, and
 $\delta m:=m_{\mu}-m_e$.

We thus see that flavor  mixing can be
seen as an interaction of the flavor fields
with an $SU(2)$ constant gauge field
having the following structure:
\bea \label{Conn}
A_{\mu} &:=&\frac{1}{2} A_{\mu}^a  \si_a \,=\,  n_{\mu} \delta m \,\frac{\sigma_1}{2}\in
su(2), \qquad n^{\mu}:=(1,0,0,0)^T ,\eea
that is, having only the temporal component in spacetime and only the
first component in $su(2)$ space. In terms of this connection, the
covariant derivative can be written in the form:
\bea D_{\mu}= \partial_{\mu} + i\, g\, \beta\, A_{\mu}, \eea
where we have defined $g:=\tan 2\theta$  as the coupling constant for
the mixing interaction. Note that in the case of maximal mixing
($\theta=\pi/4$), the coupling constant grows to infinity while
$\delta m$ goes to zero.  We further note that, since  the gauge
connection is  a constant, with just one non-zero component in
group space, its field strength vanishes identically:
\bea F_{\mu\nu}^a=\epsilon^{abc}A_{\mu}^b A_{\nu}^c=0,\eea
with $a,b,c=1,2,3$. The fact that, despite $F_{\mu\nu}$ vanishes identically, the gauge field has physical effects, leads to an analogy with the Aharonov--Bohm effect \cite{Aharonov:1959fk}.

Finally, the equations of motion for the
mixed fields can be cast in a manifestly covariant form:
\bea (i\gamma^{\mu}D_{\mu} - M_d) \nu_f =0, \eea
and the Lagrangian density (\ref{Lagrflav}) has the form of the one
describing a doublet of Dirac fields in interaction with an
external Yang-Mills field:
\bea \label{LagrflavCov}{\cal L} = {\bar \nu}_f
(i\gamma^{\mu}D_{\mu} - M_d) \nu_f. \eea

\subsection{Energy-momentum tensor and 4-momentum operator}

In this Section we study the
energy momentum tensor associated with the flavor neutrino fields
in interaction with the external gauge field. This object can be computed
by means of the standard procedure \cite{LeiteLopes:1981fr}.
One finds:
\bea \label{newemtensor}\widetilde{T}_{\rho\sigma} = {\bar \nu}_f i
\gamma_{\rho}D_{\sigma}\nu_f - \eta _{\rho\sigma}{\bar \nu}_f
(i\gamma^{\lambda}D_{\lambda} - M_d)\nu_f.\eea

This expression is to be compared with the one of the canonical energy
momentum tensor given in Eq.(\ref{Temu})
from which we see that the difference between the two is just the presence of
the interaction terms in the $00$ component, i.e. $T_{00}-\widetilde{T}_{00} =
m_{e\mu}({\bar \nu}_e \nu_{\mu} +  {\bar
\nu}_{\mu} \nu_e) $, while we have $T_{0i}=\widetilde{T}_{0i}$, $T_{ij}=\widetilde{T}_{ij}$.

The tensor
$\widetilde{T}_{\mu\nu}$ is not conserved on-shell. In particular we have:
\bea \label{noncoserv}
&&
\partial^{\rho}\widetilde{T}_{\rho i}\,=\,0;
\qquad \quad
\partial^{\rho}\widetilde{T}_{\rho 0}\neq 0.
\eea
Note that without the $\beta$ matrix appearing in the
covariant derivative (\ref{covdevmix}) we would have found:
$\partial^{\mu}\widetilde{T}_{\mu\nu}= g F_{\mu\nu a}
j^{\mu}_a\,=\,0,$
i.e. the energy-momentum tensor would have been conserved. In the present case $[\gamma_\mu,D_0]\neq 0$,
in consequence of the presence of the $\beta$ matrix in $D_0$.

We also note that the matter current  $j^{\mu}_a$
has only one component in
group space:
\bea j^{\rho}_1= {\bar
\nu_f}\gamma^{\rho}\frac{\sigma_1}{2}\nu_f=\frac{1}{2}({\bar
\nu_e}\gamma^{\rho}\nu_{\mu} + {\bar \nu_{\mu}}\gamma^{\rho}\nu_e) =
J^{\rho}_{f,1}, \eea
where $J^{\rho}_{f,1}$ is the Noether current associated to the
Lagrangian density (\ref{LagrflavCov}) under the $SU(2)$
transformation \cite{BJV01}:
\bea \nu_f'= e^{i m_{e\mu} \lambda_1 \frac{\sigma_1}{2}}\nu_f.\eea

Following the usual procedure, we now define a $4$-momentum
operator $\widetilde{P}^{\mu}$ by taking the
$0i$ and $00$ components of $\widetilde{T}^{\mu\nu}$ and integrating them over
$3-$space. We obtain a conserved $3-$momentum operator:
\bea \non
\widetilde{P}^i &=& \int d^3\mathbf{x} \,\widetilde{T}^{0i}= i \int d^3\mathbf{x}\,
\nu_f^{\dagger}\partial^i \nu_f
\\ \non
&=& i \int d^3\mathbf{x}\,
\nu_e^{\dagger}\partial^i \nu_e + i \int d^3\mathbf{x}\,
\nu_{\mu}^{\dagger}\partial^i \nu_{\mu}
\\  \label{Pgauge} &\equiv & \widetilde{P}^i_e(x_0) +
\widetilde{P}^i_{\mu}(x_0),
\qquad\qquad\qquad\qquad i=1,2,3 \eea
and a non conserved Hamiltonian operator:
\bea \non \widetilde{P}^0(x_0) \equiv \widetilde{H}(x_0) &=& \int d^3 \mathbf{x}\,
\widetilde{T}^{00}=  \int d^3 \mathbf{x} \,{\bar \nu}_f\lf( i  \gamma_0 D_0
- i \gamma^\mu D_\mu + M_d \ri)\nu_f
\\ \non
&=& \int d^3 \mathbf{x}\, \nu_e^{\dagger}\lf(-i\vect{\al}\cdot\vect{\nabla}+\beta
m_e\ri) \nu_e + \int d^3 \mathbf{x}
\,\nu_{\mu}^{\dagger}\lf(-i\vect{\al}\cdot\vect{\nabla}+\beta m_{\mu}\ri)
\nu_{\mu}
\\  \label{Hgauge}
&\equiv &\widetilde{H}_e(x_0) + \widetilde{H}_{\mu}(x_0).\eea
We see that both the Hamiltonian and the momentum  operators split
in a natural way in a contribution involving only the electron
neutrino field and in another where only the muon neutrino field
appears. In such a way, we have a natural definition of a
Hamiltonian and momentum operators for each flavor field.

We remark that the tilde Hamiltonian is \textit{not} the generator of time translations. This role competes to the complete Hamiltonian $H=\int d^3\mathbf{x}\,T^{00}$, which includes the interaction term.

\section{Flavor neutrino states and Lorentz invariance}

Till now our considerations have been purely classical. Now we
want to pass to the quantum theory. Our purpose is to construct
flavor neutrino states which are simultaneous eigenstates of the
$4-$momentum operators above constructed and of the flavor
charges. Of course this is a highly nontrivial request. We will
see that such states can indeed be constructed, but this involves
a nontrivial redefinition of the flavor vacuum which will also
erase any reference to the $\nu_1$ and $\nu_2$ fields.

As  shown in Appendix A, the flavor neutrino field
operators can be expanded in the same basis as the free fields
with masses $m_1$ and $m_2$:
\bea \label{flavexp1}\nu_{\sigma}(x) = \int \frac{d^3
\mathbf{k}}{(2\pi)^{3/2}}\sum_r\lf[u^r_{\mathbf{k},j}(x_0)
\alpha^r_{\mathbf{k},\sigma}(x_0) +
v^r_{-\mathbf{k},j}(x_0)\beta^{r \dagger}_{\mathbf{-k},\sigma}(x_0)\ri]
e^{i\mathbf{k}\cdot\mathbf{x}},
\qquad (\sigma, j)=(e,1)(\mu,2), \eea
where $\alpha_{\mathbf{k},\sigma}$ and $\beta_{\mathbf{-k},\sigma}$ are the flavor ladder operators
\cite{BV95}. In the same Appendix we show that flavor neutrino states, defined as $|\nu^r_{\mathbf{k},\sigma}\rangle=
\alpha^{r \dag}_{\mathbf{k},\sigma}|0\rangle_{e,\mu}$, are eigenstates of the flavor charge operators $Q_{\sigma}$, at a given time.
They turn out also to be eigenstates of the momentum operators $P^i=\int d^3\mathbf{x}\,T^{0i}$.
However, since the Hamiltonian operator $H$ does not commute with the charges
$Q_{\sigma}$, the above flavor states do not have definite energies.

We will now show that this problem can be solved by noting that
the expansion (\ref{flavexp1}) actually relies on a special choice
of the bases of spinors, namely those referring to the free field
masses $m_1$, $m_2$. It is however always possible to perform a
Bogoliubov transformation in order to expand the field operators
in a different basis of spinors, referring to an arbitrarily
chosen couple of mass parameters \cite{Fujii:1999xa}.
Let $\mu_\sigma$, $\sigma =e,\mu$ be such a couple of arbitrary parameters.

The Bogoliubov transformation to be performed is the
following:
\bea \label{fujiiBog}
\lf( \ba{c} \widetilde{\alpha}_{\mathbf{k},\sigma}^r(x_0)\\
\widetilde{\beta}_{-\mathbf{k},\sigma}^{r\dagger}(x_0)\ea\ri) =
J^{-1}_{\mu}(x_0) \lf( \ba{c} \alpha_{\mathbf{k},\sigma}^r(x_0)\\
\beta_{-\mathbf{k},\sigma}^{r\dagger}(x_0)\ea\ri) J_{\mu}(x_0),\eea
where the generator is \cite{Fujii:1999xa}:
\bea\label{NewGen} J_{\mu}(x_0)=\prod_{\mathbf{k},r}\exp\left\{
i\sum_{(\sigma,
j)}\xi^{\mathbf{k}}_{\sigma,j}\lf[\alpha_{\mathbf{k},\sigma}^{r \dagger }(x_0)
\beta_{-\mathbf{k},\sigma}^{r \dagger}(x_0) +
\beta_{-\mathbf{k},\sigma}^r(x_0)\alpha_{\mathbf{k},\sigma}^r(x_0)\ri]
\right\}, \qquad (\sigma,j)=(e,1),(\mu,2),\eea
and $\xi^{\mathbf{k}}_{\sigma,j}=(\chi_{\sigma}-\chi_j)/2$ and
$\chi_{\sigma}= \arctan(\mu_{\sigma}/ |\mathbf{k}|)$, $\chi_j=
\arctan(m_j/ |\mathbf{k}|)$.

The explicit form of the
transformation (\ref{fujiiBog}) is the following:
\bea \lf( \ba{c} \widetilde{\alpha}_{\mathbf{k},\sigma}^r(x_0)\\
\widetilde{\beta}_{-\mathbf{k},\sigma}^{r\dagger}(x_0)\ea\ri) = \lf( \ba{cc}\rho_{\sigma, j}^{\mathbf{k}}& i\lambda_{\sigma, j}^{\mathbf{k}} \\ i\lambda_{\sigma, j}^{\mathbf{k}} & \rho_{\sigma, j}^{\mathbf{k}}\ea\ri)\lf( \ba{c} \alpha_{\mathbf{k},\sigma}^r(x_0)\\
\beta_{-\mathbf{k},\sigma}^{r\dagger}(x_0)\ea\ri),
\qquad (\sigma,j)=(e,1),(\mu,2),\eea
where $ \rho_{\sigma, j}^{\mathbf{k}}=\cos\xi^{\mathbf{k}}_{\sigma,j}$ and $\lambda_{\sigma, j}^{\mathbf{k}}=\sin\xi^{\mathbf{k}}_{\sigma,j}$.

We have thus a whole family of flavor
vacua, denoted with a tilde and parameterized by the couples $(\mu_e,\mu_{\mu})$:
\bea |{\widetilde 0}(x_0)\rangle_{e\mu}=J_{\mu}^{-1}(x_0)|
0(x_0)\rangle_{e\mu}.\eea
The original flavor vacuum is of course the one associated
with the couple $(m_1, m_2)$.

Notice that the
flavor charges, as well as the exact oscillation formulae are
invariant under the above Bogoliubov transformations \cite{remarks}, i.e. $\widetilde{Q}_{\sigma}=Q_{\sigma}$, with:
\bea
\widetilde{Q}_{\sigma}(x_0)=\sum_r \int
d^3\mathbf{k}\lf(\widetilde{\alpha}^{r \dagger}_{\mathbf{k}\sigma}(x_0)
\widetilde{\alpha}^r_{\mathbf{k}\sigma}
(x_0)- \widetilde{\beta}^{r \dagger}_{-\mathbf{k}\sigma}(x_0)
\widetilde{\beta}^r_{-\mathbf{k}\sigma}(x_0)\ri).\eea

In the context of the above reformulation of mixing as a gauge theory, it
seems natural to expand the flavor fields in the bases
corresponding to the couple of masses
$(m_e,m_{\mu})$. We will discover that precisely those values are
singled out by the requirement that the flavor states be
eigenstates of the Hamiltonian operator.

The new spinors are defined as the solutions of the
equations:
\bea (-\alpha\cdot \mathbf{k} + m_\si \beta)
u^r_{\mathbf{k},\si} &=& \omega_{\mathbf{k},\si} u^r_{\mathbf{k},\si}
\\ [2mm]
(-\alpha\cdot\mathbf{k} + m_\si \beta) v^r_{-\mathbf{k},\si} &=&
-\omega_{\mathbf{k},\si} v^r_{-\mathbf{k},\si}, \eea
where $\omega_{\mathbf{k},\si}= \sqrt{\mathbf{k}^2+m_\si^2}$. These
are the momentum space version of the \emph{free} Dirac equation
with mass $m_\si$.

The flavor field operators are
then expanded as follows:
\bea \nu_\si(x) = \int \frac{d^3
\mathbf{k}}{(2\pi)^{3/2}}\sum_r\lf[u^r_{\mathbf{k},\si}(x_0)
\widetilde{\alpha}^r_{\mathbf{k},\si}(x_0)
+ v^r_{-\mathbf{k},\si}(x_0)
\widetilde{\beta}^{r \dagger}_{\mathbf{-k},\si}(x_0)\ri]
e^{i\mathbf{k}\cdot\mathbf{x}}, \quad \sigma=e,\mu, \eea
with $u^r_{\mathbf{k},\si}(x_0) = u^r_{\mathbf{k},\si} e^{-i
\omega_{\mathbf{k},\si} x_0}$,
$v^r_{-\mathbf{k},\si}(x_0)=v^r_{-\mathbf{k},\si}e^{i
\omega_{\mathbf{k},\si} x_0}$.
Here and in the following the tilde operators are those corresponding
to the specific couple
$(m_e,m_{\mu})$.
With these definitions all the calculations at a fixed instant of
time $x_0$ can be performed in exactly the same way they are done
in the free field case. The explicit time dependence of the
creation and destruction operators, which is of course due to the
interaction with the external field and is not present in the free
field case, does not create problems as the states which are acted
upon by the operators are evaluated at the same time as the
operators themselves and the commutators are all considered at
equal times.

In terms of the tilde flavor ladder operators, the Hamiltonian
and momentum operators Eqs.(\ref{Pgauge}),(\ref{Hgauge}) read:
\bea \widetilde{\mathbf{P}}_\si(x_0)
&=& \sum_r\int d^3 {\bf k}\,\,  {\bf k} \left(
\widetilde{\alpha}^{r\dagger}_{{\bf k},\si}(x_0) \widetilde{\alpha}^{r}_{{\bf
k},\si}(x_0) + \widetilde{\beta}^{r\dagger}_{{\bf {k}},\si}(x_0)
\widetilde{\beta}^{r}_{{\bf k},\si}(x_0) \right),
\\ [2mm]
\widetilde{H}_{\sigma}(x_0)&=&
\sum_r\int d^3{\bf k}\,  \om_{\mathbf{k},\si} \left(
\widetilde{\alpha}^{r\dagger}_{{\bf k},\si}(x_0)
\, \widetilde{\alpha}^{r}_{{\bf k},\si} (x_0)-
\widetilde{\beta}^{r}_{{\bf {k}},\si} (x_0)
\, \widetilde{\beta}^{r\dagger}_{{\bf k},\si} (x_0)\right)
. \eea

The new flavor states are defined by the action of the tilde
creation operator on the tilde flavor vacuum:
\bea \label{NewFlavState}
|\widetilde{\nu}^{\, r}_{\mathbf{k},\sigma}(x_0)\rangle
= \widetilde{\alpha}^{r \dagger}_{\mathbf{k},\sigma}(x_0)|\widetilde{0}(x_0)\rangle_{e\mu}.\eea
We easily find the result that these single particle states are
eigenstates of both the Hamiltonian and the momentum operator:
\bea
\lf(\ba{c}
\widetilde{H}_{\sigma}(x_0)\\\widetilde{\mathbf{P}}_{\sigma}(x_0)\ea\ri)
|\widetilde{\nu}^{\, r}_{\mathbf{k},\sigma}(x_0)\rangle
= \lf(\ba{c}
\omega_{\mathbf{k},\sigma}\\
\mathbf{k}\ea\ri)|\widetilde{\nu}^{\, r}_{\mathbf{k},\sigma}(x_0)\rangle,
\eea
making explicit the $4-$vector
structure.

It can be also verified that the flavor charges
commute with the tilde Hamiltonian operator: $[\widetilde{Q}_{\sigma}(x_0),\widetilde{H}(x_0)]=0$, as a consequence of:
\bea
[\widetilde{Q}_{\sigma}(x_0),\widetilde{H}_{\sigma'}(x_0)]=0,\qquad
\sigma,\sigma'=e,\mu.
\eea
This is of course a consequence of the fact that the flavor
nonconservation is entirely due to the interaction term, which is
absent in $\widetilde{H}$. This fact ensures the existence of a
common set of eigenstates of these operators. Indeed the flavor
states (\ref{NewFlavState}) are straightforwardly seen to be also
eigenstates of the flavor charges:
\bea
\widetilde{Q}_\si(x_0)|\widetilde{\nu}^{\, r}_{\mathbf{k},\si}(x_0)\rangle=
|\widetilde{\nu}^{\, r}_{\mathbf{k},\si}(x_0)\rangle, \eea
thus confirming that these are precisely the states we were looking
for.

\medskip

We would like to conclude this Section by making some observations on the
algebra of the generators descending from the energy-momentum
tensor (\ref{newemtensor}). All the generators are defined in
the usual way. Besides the translation generators defined by
Eqs.(\ref{Pgauge}) and (\ref{Hgauge}), we have the Lorentz
generators, defined as:
\bea \widetilde{M}^{\lambda\rho}(x_0)=\int
d^3\mathbf{x}\lf(\widetilde{T}^{0\rho}x^{\lambda}-
\widetilde{T}^{0\lambda}x^{\rho}\ri)
+ \frac{1}{2}\int
d^3\mathbf{x}\,\nu_f^{\dagger}\sigma^{\lambda\rho}\nu_f =
\widetilde{M}^{\lambda\rho}_e(x_0)+\widetilde{M}^{\lambda\rho}_{\mu}(x_0),
\eea
where $\sigma^{\mu\nu}=-\frac{i}{2}[\gamma^{\mu},\gamma^{\nu}]$.
The algebra of
(equal-time) commutators of these generators will be just the
direct sum of two Poincar\'{e} algebras (we omit the specification
of the instant of time):
\bea \non &&[\widetilde{P}_{\si}^\mu, \widetilde{P}_{\si'}^\nu] \,=\,0
\quad;\qquad
[\widetilde{M}_{\si}^{\mu\nu},\widetilde{P}_{\si'}^\la] \,=\, i
\de_{\si\si'}\, \lf( \eta^{\mu\la} \widetilde{P}_{\si}^\nu \, - \,
\eta^{\nu\la} \widetilde{P}_{\si}^\mu \ri);
\\ [2mm]
&&[\widetilde{M}_{\si}^{\mu\nu},\widetilde{M}_{\si'}^{\la\rho}]
\,=\,i \de_{\si\si'}\, \lf( \eta^{\mu\la}
\widetilde{M}_{\si}^{\nu\rho} \, - \, \eta^{\nu\la}
\widetilde{M}_{\si}^{\mu\rho} \, - \, \eta^{\mu\rho}
\widetilde{M}_{\si}^{\nu\la} \, + \, \eta^{\nu\rho}
\widetilde{M}_{\si}^{\mu\la}\ri),\qquad
\sigma,\sigma'=e,\mu. \eea

Note that the above construction and the consequent Poincar\'e invariance,
holds \textit{at a given time} $x_0$. Thus, for each different time, we have a different  Poincar\'e structure.

\subsection{Phenomenological consequences}

The above analysis leads us to the view that the flavor fields $\nu_e$ and $\nu_\mu$
should be regarded as fundamental. This fact has some interesting consequences at phenomenological level. Indeed, if we consider a charged current process in which
for example an
electron neutrino is created,
we see that the hypothesis that mixing is due to interaction with an
external field, implies that what is created in the vertex is really $|\nu_e\ran$,
rather than $|\nu_1\ran$ or  $|\nu_2\ran$. As remarked above, such an interpretation
is made possible because we can regard, at any given time, flavor fields as on shell fields, associated
with masses $m_e$ and $m_\mu$.

We consider the case of a beta decay process, say for definiteness tritium decay, which allows for
a direct investigation of neutrino mass. In the following we compare
the various possible outcomes of this experiment predicted by the
different theoretical possibilities for the nature of mixed neutrinos.
As we shall see, the scenario described above presents significative phenomenological differences with respect to the standard theory.

Let us then consider the decay:
\bea\non A\rightarrow B + e^- +{\bar \nu}_e, \eea
where $A$ and $B$ are two nuclei (e.g. $^3$H and $^3$He).

The electron  spectrum is proportional to phase volume factor $E p
E_{e} p_{e}$:
\bea \label{specpref} \frac {dN}{dK}= C E p\,
(Q-K)\sqrt {(Q-K)^2\,-\,m_{\nu}^2} \eea
where $E=m+K$  and $p=\sqrt{E^2-m^2}$  are  electron's
energy and momentum. The endpoint of $\beta$ decay is the  maximal kinetic energy $K_{max}$
the electron can take (constrained by the available energy
$Q=E_A-E_B-m\approx m_A-m_B-m$). In the case of  tritium decay, $Q=18.6$ KeV.
$Q$ is shared between the
(unmeasured) neutrino energy and the (measured) electron kinetic
energy $K$.It is clear that if  the neutrino were massless, then $m_\nu=0$ and $K_{max}=Q$.
On the other hand, if the neutrino were a mass eigenstate with $m_\nu=m_1$,
then $K_{max}=Q-m_1$.

We now consider the various possibilities which can arise
in the presence of  mixing. If, following the common wisdom, neutrinos with masses $m_1$ and $m_2$ are considered as
fundamental, the $\beta$  spectrum is:
\bea \label{specmass}
\frac {dN}{dK}= C E p \,E_e \sum_j |U_{ej}|^2 \sqrt{E_e^2-m_{j}^2}\;
\Theta (E_e-m_j),
\eea
where $E_e=Q-K$ and
$U_{ej}=(\cos\theta,\sin\theta)$ and $\Theta (E_e-m_j)$ is the Heaviside step function.
The end point is at $K=Q-m_1$  and the spectrum has an
inflexion at $K\simeq Q-m_2$.

\begin{figure}
\begin{center}
\includegraphics*[width=11.5cm]{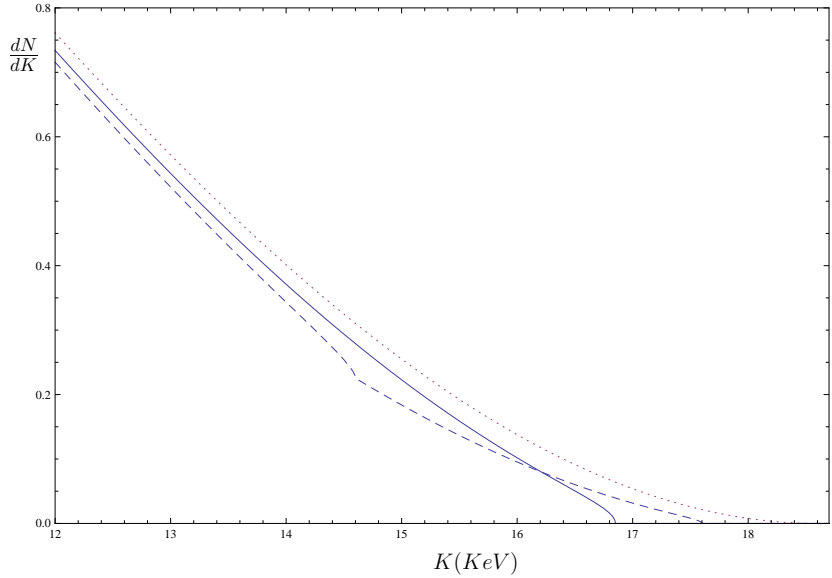}
\vspace{-0.5cm}
\end{center}
\caption{The tail of the tritium $\beta$ spectrum for:
- a massless neutrino (dotted line);
- fundamental flavor states (continuous line);
- superposed prediction for 2 mass states (short-dashed line):
notice the inflexion in the spectrum where the most massive state
switches off.
We used $m_e=1.75$ KeV, $m_1=1$ KeV, $m_2=4$
KeV, $\theta=\pi/6$.} \label{fig1}
\end{figure}

If on the other hand we take  flavor neutrinos as fundamental according to the above scheme, we have that  $m_\nu=m_e$ and $K_{max}=Q- m_{e}$   and the spectrum is proportional to
the phase volume factor $E p E_{e}p_{e}$:
\bea
\frac {dN}{dK}= C E p\, (Q-K)\sqrt {(Q-K)^2\,-\,m_{e}^2}\; \Theta
(E_e-m_e).\label{specpref}
\eea
%

The above discussed possibilities are plotted in Fig.(\ref{fig1}), together
with the spectrum for a massless neutrino, for comparison.
We note that the next generation tritium beta decay experiments will allow a sub-eV
sensitivity for the electron neutrino mass \cite{katrin}, thus hopefully allowing to
unveil the true nature of mixed neutrinos.

\medskip

Finally we point out that also in the neutrino detection
process, it would be possible
to discriminate among the various scenarios above considered.
In such a case, our scheme
would imply that in each detection vertex,
either an electron neutrino or a muon neutrino
would take part to the process.
Again, this is in contrast with the standard view, which
assumes that either $\nu_1$ or $\nu_2$ are
entering in the elementary processes.

\section{Discussion and Conclusions}

In this paper we have studied a non-abelian
gauge structure associated to flavor mixing. In this framework flavor neutrino fields are taken to be on-shell fields with definite masses $m_e$ and $m_{\mu}$, which
are different from those of the  mass eigenstates of the standard approach, $m_1$ and $m_2$. Flavor  oscillations then arise as a consequence of the interaction with the gauge field, which acts as a  sort of refractive medium --  \emph{neutrino aether}.

It would be interesting to explore the properties of such a medium and possible optical analogs of this situation. A very interesting example in this respect
has been given recently in Ref.\cite{Weinheimer:2010ar}.
Another interesting analogy can be drawn with recent studies in which, for the case of photons, the vacuum  has been  thought to act as a refractive medium in  consequence of quantum gravity fluctuations \cite{Ellis:2008gg}.

The gauge structure associated to flavor mixing
has the very interesting property of arising
across different fermion generations,
thus having a different (``horizontal'') nature
with respect the  gauge structure of the Standard Model.

A  natural question that comes in concerns the origin of the external gauge field which causes the mixing. This  could also have  some connection to quantum gravity models that sometimes are invoked to explain the origin of mixing \cite{Mavromatos:2006ux}.

Another outcome of our analysis is that we could recover, at least locally in time, a Poincar\'e structure for the flavor states. This is possible since we could define a Hamiltonian operator that commutes with the flavor charges, thus allowing for simultaneous eigenstates. In this scheme where the fields $\nu_e$ and $\nu_\mu$ are taken to be fundamental, one avoids any reference to the fields $\nu_1$ and $\nu_2$.  As pointed out, this leads to phenomenological consequences
that can be possibly tested in experiments on beta decay.

\medskip

A final consideration concerns the interpretation of the
Hamiltonian operator $\widetilde{H}$ which, as already remarked, does not
take into account the interaction energy, i.e. the energy
associated with mixing. We can thus view $\widetilde{H}$ as the sum of the
kinetic energies of the flavor neutrinos, or equivalently as the energy which can be extracted from flavored
neutrinos by scattering processes, the mixing energy being
``frozen'' (there's no way to turn off the mixing!). This suggests the interpretation of such a quantity as a ``free''
energy $F\equiv \widetilde{H} $, so that we can write:
\bea H-F= T S. \eea
This quantity defines an entropy associated with flavor mixing. It is natural
to identify the ``temperature'' $T$ with the coupling constant $g=\tan 2\theta$, thus leading to:
\bea \label{entropy}
 S &=&  \int  d^3\mathbf{x}\,
{\bar \nu}_f A_0 \nu_{f}\,=\,\frac{1}{2}\, \delta m \int  d^3\mathbf{x}\,
({\bar \nu}_e \nu_{\mu} +  {\bar
\nu}_{\mu} \nu_e).
\eea

The appearance of an entropy should not be surprising, since each of the two flavor neutrinos can be considered as an open system which presents some kind of (cyclic) dissipation. This situation can be handled by use of well known methods of Thermo Field Dynamics \cite{Umezawa:1982nv} developed for the study of quantum dissipative systems \cite{CRV92}.

The explicit expression for the expectation values of the entropy on the flavor neutrino states is quite complicated, and thus not very illuminating. An attempt at an interpretation of it is given in Appendix B in the much simpler context of Quantum Mechanics. There  it is shown that  at a given time, the difference of the expectation values of the muon and electron free energies is less than the total initial energy of the flavor neutrino state. The missing part is proportional to the expectation value of the entropy.

The scenario emerged in this work, and in particular the last considerations, is consistent with an interpretation of the gauge field as a reservoir, first put forward in \cite{Celeghini:1992ea}.

Finally, we point out that recent work \cite{Blasone:2007wp,Blasone:2007vw,Blasone:2010ta} has shown that a time dependent entanglement entropy is associated with neutrino mixing and oscillations. It is an interesting question the one of the connection of the latter to the above entropy.

\section*{Acknowledgements}

Support from
INFN and MIUR is partially acknowledged.

\appendix
\section{Mixing of quantum neutrino fields}

In this appendix we briefly recall the quantization for mixed fields, as given in
Refs.\cite{BV95,BHV98,BJV01}.
We start from the  free fields $\nu_1$ and $\nu_2$, whose Fourier expansions are:
\bea \nu_j(x) = \int \frac{d^3
\mathbf{k}}{(2\pi)^{3/2}}\sum_r\lf[u^r_{\mathbf{k},j}(x_0)\alpha^r_{\mathbf{k},j}
+ v^r_{-\mathbf{k},j}(x_0)\beta^{r \dagger}_{\mathbf{-k},j}\ri]
e^{i\mathbf{k}\cdot\mathbf{x}}, \qquad j=1,2,
\eea
with $u^r_{\mathbf{k},j}(x_0) = u^r_{\mathbf{k},j} e^{-i
\omega_{\mathbf{k},j} x_0}$,
$v^r_{-\mathbf{k},j}(x_0)=v^r_{-\mathbf{k},j}e^{i
\omega_{\mathbf{k},j} x_0}$, and
$\omega_{\mathbf{k},j}=\sqrt{\mathbf{k}^2+m_j^2}$. The operators
$\alpha^r_{\mathbf{k},j}$ and $\beta^{r}_{\mathbf{-k},j}$,
$j=1,2$, $r=1,2$ are the annihilation operators for the vacuum
state $|0\rangle_{1,2}=|0\rangle_1\otimes|0\rangle_2$:
$\alpha^r_{\mathbf{k},j}|0\rangle_{1,2},
\beta^{r}_{\mathbf{-k},j}|0\rangle_{1,2}=0$. The canonical
anticommutation relations are: $\{\nu_i^{\alpha}(x),
\nu_j^{\beta\dagger}(y)\}_{x_0=y_0}=\delta^3(\mathbf{x}-\mathbf{y})
\delta_{\alpha\beta}\delta_{ij}$
with $\alpha, \beta=1,\ldots,4$ and
$\{\alpha^r_{\mathbf{k},i},\alpha^{s\dagger}_{\mathbf{q},j}
\}=\delta_{\mathbf{k}\mathbf{q}}\delta_{rs}\delta_{ij}$; $\{
\beta^r_{\mathbf{k},i},\beta^{s\dagger}_{\mathbf{q},j}
\}=\delta_{\mathbf{k}\mathbf{q}}\delta_{rs}\delta_{ij}$, with
$i,j=1,2$. All other anticommutators are zero. The ortonormality
and completeness relations are: $u_{{\bf k},j}^{r\dagger }u_{{\bf
k},j}^{s} = v_{{\bf k},j}^{r\dagger }v_{{\bf k},j}^{s}= \delta
_{rs}$, $u_{{\bf k},j}^{r\dagger }v_{-{\bf k},j}^{s} = v_{-{\bf k}
,j}^{r\dagger }u_{{\bf k},j}^{s}=0$, $ \sum_{r}(u_{{\bf
k},j}^{r}u_{{\bf k},j}^{r\dagger }+v_{-{\bf k},j}^{r}v_{-{\bf
k},j}^{r\dagger }) = 1$.

We construct the generator for the mixing transformations
(\ref{PontecorvoMix}) as:
\bea \label{MixingRel} \nu^{\alpha}_\si(x)&=&
G_{\theta}^{-1}(x_0)\nu^{\alpha}_j G_{\theta}(x_0) \,,
\qquad (\sigma, j)=(e,1),(\mu,2), \eea
with $G_{\theta}(x_0)$  given by:
\be \label{MixGen} G_{\theta}(x_0)=\exp\left[\theta\int
d^3\mathbf{x}\left(\nu_1^{\dagger}(x)\nu_2(x)-
\nu_2^{\dagger}(x)\nu_1(x)\right)\right], \ee
At finite volume $G_{\theta}$ is a unitary operator:
$G_{\theta}^{-1}(x_0)=G_{-\theta}(x_0)=G_{\theta}^{\dagger}(x_0)$,
preserving the canonical anticommutation relations. $G_{\theta}^{-1}(x_0)$ maps the Hilbert space for the free fields
$\mathcal{H}_{1,2}$ to the Hilbert space for the mixed fields
$\mathcal{H}_{e,\mu}$:
$G_{\theta}^{-1}(x_0):\mathcal{H}_{1,2}\rightarrow\mathcal{H}_{e,\mu}$.
In particular for the vacuum $|0\rangle_{1,2}$ we have, at finite
volume $V$:
\be\label{flavvac} |0(x_0) \rangle_{e,\mu} = G^{-1}_{\bf \te}(x_0)\;
|0 \rangle_{1,2}\;; \ee
$|0 \rangle_{e,\mu}$ is the vacuum for the Hilbert space
$\mathcal{H}_{e,\mu}$, which we will refer us to as the flavor
vacuum.

In the limit $V\rightarrow\infty$ the flavor vacuum becomes
orthogonal to the vacuum of the free fields, which means that the
two Hilbert spaces are unitarily inequivalent. Due to the
linearity of $G_{\theta}(x_0)$, we can define the flavor
annihilators as:
\begin{eqnarray}\label{flavannich}
\alpha _{{\bf k},\sigma}^{r}(x_0) &\equiv &G^{-1}_{\bf
\te}(x_0)\;\alpha
_{{\bf k},j}^{r}\;G_{\bf \te}(x_0),   \qquad (\sigma, j)=(e,1), (\mu,2)
\end{eqnarray}
and similar ones for the antiparticle operators.
In the reference frame for which $\mathbf{k}=(0,0,|\mathbf{k}|)$, the electron neutrino annihilation operator has the form:
\bea \alpha^{r}_{{\bf k},e}(x_0)&=&\cos\theta\;\alpha^{r}_{{\bf
k},1}\;+\;\sin\theta\;\left( U_{{\bf k}}(x_0)\; \alpha^{r}_{{\bf
k},2}\;+\;\epsilon^{r}\; V_{{\bf k}}(x_0)\; \beta^{r\dag}_{-{\bf
k},2}\right),
\eea
and similar expressions hold for the other ladder operators \cite{BV95}.
Here $\epsilon^{r}=(-1)^{r}$ and:
\bea\label{Vk2}
 U_{{\bf k}}(x_0)&\equiv &u^{r\dag}_{{\bf
k},2}(x_0)u^{r}_{{\bf k},1}(x_0)= v^{r\dag}_{-{\bf
k},1}(x_0)v^{r}_{-{\bf k},2}(x_0);
\\
V_{{\bf k}}(x_0)&\equiv & \epsilon^{r}\; u^{r\dag}_{{\bf
k},1}(x_0)v^{r}_{-{\bf k},2}(x_0)= -\epsilon^{r}\; u^{r\dag}_{{\bf
k},2}(x_0)v^{r}_{-{\bf k},1}(x_0), \eea

and we have  $U_{k}(x_0)=|U_{k}|\;e^{i(\omega_{k,2}-\omega_{k,1})x_0}\;,
V_{k}(x_0)=|V_{k}|\;e^{i(\omega_{k,2}+\omega_{k,1})x_0}$, with $|U_{k}|^2 \;+\; |V_{k}|^2 \;=\;1$.

The flavor fields are then rewritten in the form:
\bea \nu_{\sigma}(x) = \int \frac{d^3
\mathbf{k}}{(2\pi)^{3/2}}\sum_r\lf[u^r_{\mathbf{k},j}(x_0)
\alpha^r_{\mathbf{k},\sigma}(x_0) +
v^r_{-\mathbf{k},j}(x_0)\beta^{r \dagger}_{\mathbf{-k},\sigma}(x_0)\ri]e^{i\mathbf{k}\cdot\mathbf{x}},
 \qquad (\sigma, j)=(e,1),(\mu,2),
\eea
i.e. they can be expanded in the same bases as the fields $\nu_i$.

The symmetry properties of the Lagrangian (\ref{Lagrflav}) have
been studied in Ref.~\cite{BJV01}: one has a total
conserved charge $Q$ associated with the global $U(1)$ symmetry
and time-dependent charges associated to the (broken) $SU(2)$
symmetry. Such charges are the relevant physical quantities for
the study of flavor oscillations. They are also essential in the
definition of physical flavor neutrino states, as the one
produced in beta decay, for example.

We obtain for the flavor charges Eq.(\ref{flavcharges}) the
expansion \cite{BJV01}:
\bea Q_{\sigma}(x_0) &=&
\sum_{r}\int d^3\mathbf{k} \left(
\alpha^{r\dagger}_{\mathbf{k},\sigma}(x_0)
\alpha^{r}_{\mathbf{k},\sigma}(x_0) -
\beta^{r\dagger}_{-\mathbf{k},\sigma}(x_0)
\beta^{r}_{-\mathbf{k},\sigma}(x_0)\right).\eea
%
Flavor neutrino states are defined as eigenstates of the flavor charges:
\bea Q_{\sigma}(x_0)|\nu_{\sigma}^{\mathbf{k}}(x_0)\rangle &=&
|\nu_{\sigma}^{\mathbf{k}}(x_0)\rangle , \eea
with $|\nu_{\sigma}^{\mathbf{k}}(x_0)\rangle =
\alpha^{r\dagger}_{\mathbf{k},\sigma}(x_0)|0(x_0)\rangle_{e,\mu}$,
and similar ones for antiparticles.

Note that the flavor charges do not commute with the Hamiltonian of the
system:
\bea [H, Q_{\sigma}(x_0)]\neq0, \qquad \sigma=e,\mu \eea
with the consequence that they
are not conserved by time evolution. This of course gives rise to
the oscillation phenomenon. The Hamiltonian and the flavor
charges are thus non compatible observables, with the consequence
that one cannot measure simultaneously the total energy and the
flavor of an oscillating neutrino. The flavor states are however eigenstates of
the momentum operator Eq.~(\ref{Pfree}):
\bea \mathbf{P}|\nu_{\sigma}^{\mathbf{k}}(x_0)\rangle &=&
\mathbf{k}|\nu_{\sigma}^{\mathbf{k}}(x_0)\rangle. \nonumber \eea

%

The flavor oscillation formulas are derived
by computing, in the Heisenberg representation, the
expectation value of the flavor charge operators on the flavor state.
We have
\bea
 _{e,\mu }\langle 0|\nof\wwQ_\si(x_0)\nof|0\rangle _{e,\mu } =0, \eea
where $:: ... ::\,$ denotes normal ordering with respect to
the vacuum $|0\ran_{e,\mu}$, defined in the usual way as $ :: A
::\, \equiv \,  A \, -\, {}_{e,\mu}\lan 0| A | 0 \ran_{e,\mu}\ $
for a generic operator $A$. The result is \cite{BHV98}:
\begin{eqnarray}\non \label{oscillfor1}
 {\cal Q}^{\bf k}_{{\nu_e}\rightarrow\nu_e}(x_0)& =&\langle \nu_{{\bf
k},e}^{r}|\nof\wwQ_e(x_0)\nof|\nu_{{\bf k},e}^{r}  \rangle
\\ [2mm]
&=& 1-\sin ^{2}(2\theta )\left[ \left| U_{\mathbf{k}}\right|
^{2}\sin ^{2}\left( \frac{\omega _{k,2}-\omega
_{k,1}}{2}x_0\right)
+\left| V_{\mathbf{k%
}}\right| ^{2}\sin ^{2}\left( \frac{\omega _{k,2}+\omega
_{k,1}}{2}x_0\right) \right],
\\ [2mm]
 \label{oscillfor2} {\cal Q}^{\bf k}_{{\nu_e}\rightarrow\nu_\mu}(x_0)
&=& \langle \nu_{{\bf k},e}^{r}
|\nof\wwQ_\mu(x_0)\nof|\nu_{{\bf k},e}^{r} \rangle \, = \, 1 \, -\, {\cal
Q}^{\bf k}_{{\nu_e}\rightarrow\nu_e}(x_0).
\end{eqnarray}
In the relativistic limit: $\left| \mathbf{k}\right| \gg \sqrt{m_{1}m_{2}}$, we have
$\left| U_{\mathbf{k}}\right| ^{2}\longrightarrow 1$ and $\left|
V_{\mathbf{k}}\right| ^{2}\longrightarrow 0$ and the traditional oscillation
formulas are recovered.

\section{The case of Quantum Mechanics}
In this appendix we develop a similar analysis to the one given in this paper
 to the case of mixing in Quantum Mechanics. This is useful for the
 interpretation of the results, which in this case have a much simpler form.

In a QM context,
the flavor (fermionic) annihilation operators are defined by the relations:
\bea
\alpha_e(t) &=& \cos\theta\,\alpha_1(t) + \sin\theta\,\alpha_2(t)
\\ [2mm]
\alpha_{\mu}(t) &=& -\sin\theta\,\alpha_1(t) + \cos\theta\,\alpha_2(t), \eea
where $\alpha_i(t) = e^{i\omega_i t}\alpha_i$, $i=1,2$. The flavor states are given by:
\bea |\nu_{\sigma}(t)\rangle = \alpha^{\dagger}_{\sigma}(t)|0\rangle_m, \qquad \sigma= e, \mu, \eea
where $|0\rangle_m=|0\rangle_1\otimes|0\rangle_2$ is the vacuum for the mass eigenstates. We use the notation $|\nu_{\sigma}\rangle=|\nu_{\sigma}(t=0)\rangle$. The Hamiltonian of the system is:
\bea\non H &=& \omega_e\alpha_e^{\dagger}(t)\alpha_e(t) + \omega_{\mu}\alpha_{\mu}^{\dagger}(t)\alpha_{\mu}(t)+ \omega_{e\mu}\lf[\alpha_e^{\dagger}(t)\alpha_{\mu}(t) + \alpha_{\mu}^{\dagger}(t)\alpha_e(t)\ri]
\\ &=& \omega_1\alpha_1^{\dagger}\alpha_1 + \omega_2\alpha_2^{\dagger}\alpha_2, \eea
where $\omega_e=\omega_1 \cos^2\theta + \omega_2 \sin^2\theta$, $\omega_{\mu}=\omega_1 \sin^2\theta + \omega_2 \cos^2\theta$, $\omega_{e\mu}=(\omega_2 - \omega_1)\sin\theta\cos\theta$.

In analogy with the QFT case we define the covariant derivative:
\bea
D_t = \frac{d}{dt} + i g A = \frac{d}{dt} + i\omega_{e\mu}\sigma_1, \eea
where we have $\omega_{e\mu}=\frac{1}{2}\tan2\theta\delta\omega$, $\delta\omega=\omega_{\mu}-\omega_e$, and we have defined $A := \delta\omega\frac{\sigma_1}{2}$. Using this covariant derivative the equations of motion read:
\bea D_t\,\alpha_f=-i\omega_d\,\alpha_f, \eea
where $\alpha_f=(\alpha_e,\alpha_{\mu})^T$ and $\omega_d=\textrm{diag}(\omega_e,\omega_{\mu})$. The Hamiltonian can then be written in the form:
\bea H=\alpha_f^{\dagger} \omega_d \alpha_f+ g \alpha_f^{\dagger} A \alpha_f. \eea
The diagonal part of the above expression can be readily split into separate contributions for each flavor
\bea\widetilde{H}(t)= \alpha_f^{\dagger}\omega_d\alpha_f = \omega_e\alpha_e^{\dagger}(t)\alpha_e(t) + \omega_{\mu}\alpha_{\mu}^{\dagger}(t)\alpha_{\mu}(t) = \widetilde{H}_e(t) + \widetilde{H}_{\mu}(t). \eea
Note that expectation values of the flavor number operators on the single particle flavor neutrino states at time zero give  the oscillation probabilities:
\bea \langle \nu_e(0)|N_e(t)|\nu_e(0)\rangle&=&P_{\nu_e\rightarrow\nu_e}(t)=1-\sin ^{2}2\theta \text{ }\sin ^{2}\left( \frac{\omega_2-\omega_1}{2}t\right); \\  \langle\nu_e(0)|N_{\mu}(t)|\nu_e(0)\rangle&=& P_{\nu_e\rightarrow\nu_{\mu}}(t)=\sin ^{2}2\theta \text{ }\sin ^{2}\left( \frac{\omega_2-\omega_1}{2}%
t\right).\eea
Thus we have:
\bea \langle \nu_e(0)|\widetilde{H}_e(t)|\nu_e(0)\rangle &=& \omega_e P_{\nu_e\rightarrow\nu_e}(t);\\ \langle \nu_e(0)|\widetilde{H}_{\mu}(t)|\nu_e(0)\rangle &=& \omega_{\mu} P_{\nu_e\rightarrow\nu_{\mu}}(t).\eea

\begin{figure}
\begin{center}
\includegraphics*[width=10cm]{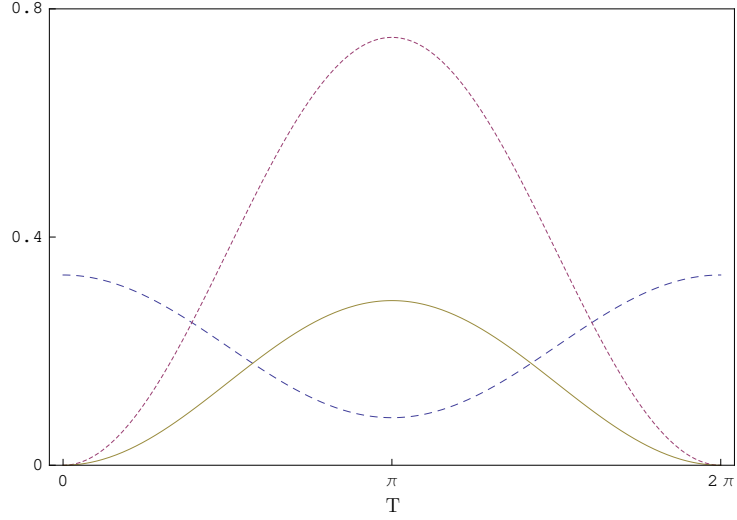}
\vspace{-0.5cm}
\end{center}
\caption{Plot of expectation values on
$|\nu_e(0)\rangle $ of $F_e(t)$ (long-dashed line),  $F_\mu(t)$ (short-dashed line) and $2 T S_e(t)$ (solid line).
 We used rescaled dimensionless time $T= (\om_2-\om_1)t$ and
$\theta=\pi/6$. The scale on the vertical axis is normalized to $\om_\mu$.} \label{fig2}
\end{figure}

In analogy with the field theoretical case, we regard these ``free'' Hamiltonians as free energies, and we write:
\bea H= \sum_{\sigma=e,\mu}(F_{\sigma}(t) + T S_{\sigma}(t)), \eea
where we make the identifications $g\equiv T$ and:
\bea
S_{\sigma}(t)=\frac{1}{4}\;\delta\omega\lf[\alpha_e^{\dagger}(t)\alpha_{\mu}(t) + \alpha_{\mu}^{\dagger}(t)\alpha_e(t)\ri]. \eea
We have:
\bea \langle \nu_e(0)|S_e(t)|\nu_e(0)\rangle = \langle \nu_e(0)|S_{\mu}(t)|\nu_e(0)\rangle= -\frac{1}{4}\;
\delta\omega\sin4\theta\sin^2
\left[\frac{1}{2}(\omega_2-\omega_1)t\right]. \eea

All the expectation values obtained are summarized in the following table,
from which we immediately see how the energetic balance is recovered. The
situation for an electron neutrino state is represented in Fig.~\ref{fig2} for
sample values of the parameters.

\begin{table}[h]
\begin{center}
\begin{tabular}{|c|c|c|c|c|}
  \hline
   & $H$ & $F_e$ & $F_{\mu}$ & $TS_e=TS_{\mu}$
  \\ \hline\hline
  $|\nu_e(0)\rangle$ & $\omega_e$ & $\omega_e(1-P(t))$ &
  $\omega_{\mu}P(t)$ & $\frac{1}{2}\delta\omega P(t)$ \\ \hline
  $|\nu_{\mu}(0)\rangle$ & $\omega_{\mu}$ & $\omega_{\mu}P(t)$ &
  $\omega_{e}(1-P(t))$ & $-\frac{1}{2}\delta\omega P(t)$ \\ \hline
\end{tabular}
\caption{Energetic balance for flavor neutrino states.  $P(t)$ denotes the transition probability $P_{\nu_e\rightarrow\nu_{\mu}}(t)$.} \label{Tab}
\end{center}
\end{table}

Note finally that the integral of the entropy expectation value over an oscillation cycle, is only dependent on the mixing angle:
\bea
\int_0^{\tau} \langle \nu_e(0)|S_e(t)|\nu_e(0)\rangle  \,dt
\,=\, \pi \cos^2 2\te\, \sin 2 \te.
\eea
where the period $\tau = \frac{2\pi}{\om_2-\om_1}$. It is interesting to
compare this result with the geometric invariants discussed  in Ref.\cite{Blasone:1999tq}.

\end{document}